\documentclass[aps,prf,preprint]{revtex4-1}

\usepackage{amssymb}
\usepackage{bm}
\usepackage{amsmath}
\usepackage{graphicx}

\begin{document}

\title{A hydrodynamic bifurcation in electroosmotically-driven periodic flows}

\author{Alexander Morozov}
\email{alexander.morozov@ph.ed.ac.uk}
\affiliation{SUPA, School of Physics and Astronomy, The University of Edinburgh, James Clerk Maxwell Building, Peter Guthrie Tait Road, Edinburgh, EH9 3FD, United Kingdom}

\author{Davide Marenduzzo}
\affiliation{SUPA, School of Physics and Astronomy, The University of Edinburgh, James Clerk Maxwell Building, Peter Guthrie Tait Road, Edinburgh, EH9 3FD, United Kingdom}

\author{Ronald G. Larson}
\affiliation{Department of Chemical Engineering, University of Michigan, Ann Arbor, MI 48109, USA}

\date{\today}

\begin{abstract}
In this paper we report a novel inertial instability that occurs in electro-osmotically driven channel flows. We 
assume that the charge motion under the influence of an externally applied electric field is confined to a small vicinity of the channel walls that, effectively, drives a bulk flow through a prescribed slip velocity at the boundaries. Here, we study spatially-periodic wall velocity modulations in a two-dimensional straight channel numerically. At low slip velocities, the bulk flow consists of a set of vortices along each wall that are left-right symmetric, while at sufficiently high slip velocities, this flow loses its stability though a supercritical bifurcation. Surprisingly, the new flow state that bifurcates from a left-right symmetric base flow has a rather strong mean component along the channel, which is similar to pressure-driven velocity profiles. The instability sets in at rather small Reynolds numbers of about 20-30, and we discuss its potential applications in microfluidic devices.
\end{abstract}

\maketitle

\begin{figure}
\centering
\includegraphics[trim={0 0 0 2.5cm},clip,width=0.6\textwidth]{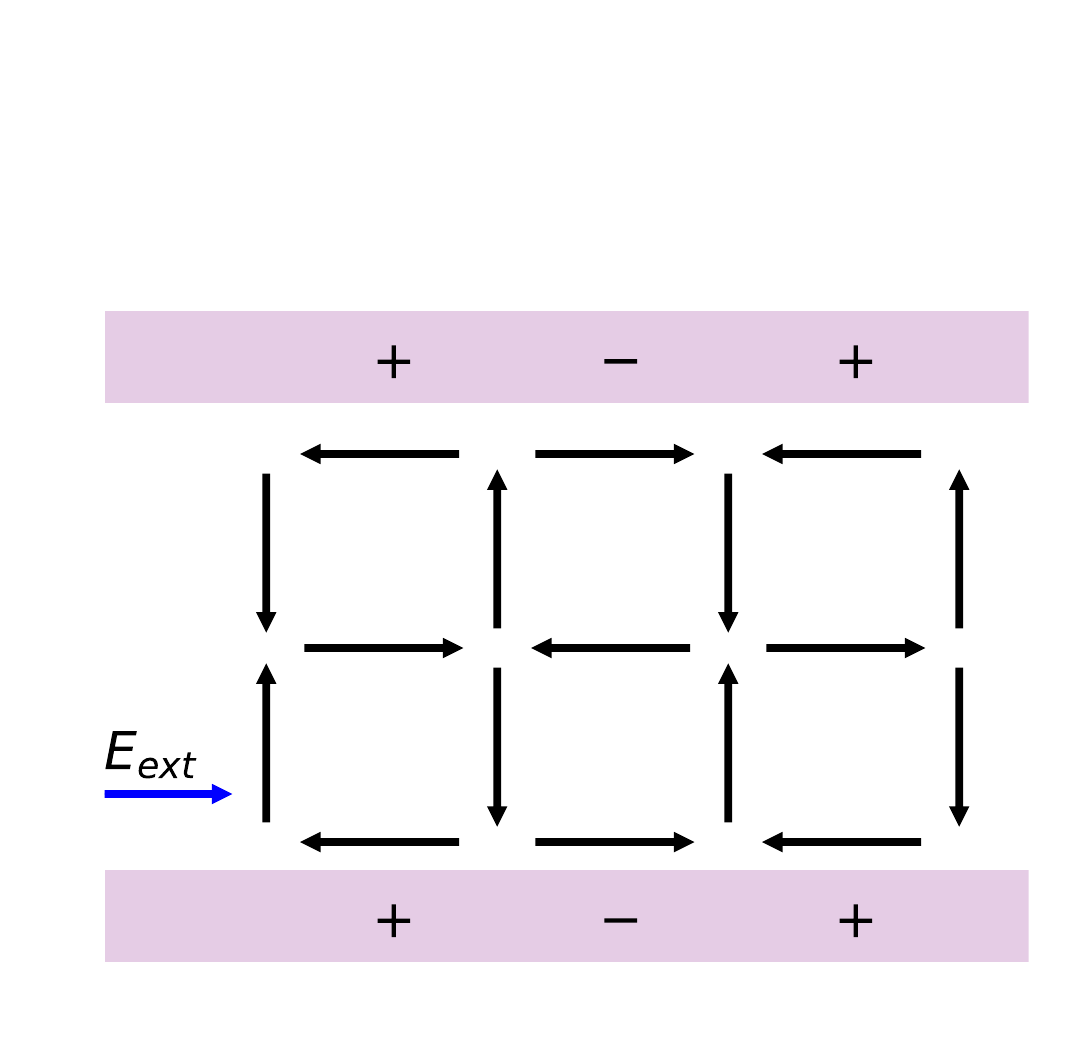}
\caption{Electro-osmotically driven periodic flow; after Ajdari \cite{Ajdari1995}. }
\label{Ajdari}
\end{figure}

In microfluidic devices, the use of electric fields as a means of driving flow via electro-osmosis is an intriguing alternative to using pressure drops or moving surfaces \cite{Pretorius1974,Probstein1994,Schoch2008,Wang2009}.
Electro-osmosis occurs when the ions in a double layer next to a charged surface are set in motion by an electric field, and the ions drag the solvent with them, producing bulk flow. Such flows are especially suitable for microfluidic applications, in which microfabrication techniques allow for control and patterning of electric and dielectric properties of channel surfaces.  In this way, not only can bulk flow be generated to transport analytes, but patterned flow fields can be imposed, allowing, for example, for creation of microfluidic mixers \cite{Lee2011}. Such flows may also assist in separating particles or cells, possibly both modulating and augmenting the inertial forces that produce size-depending cross-stream drift \cite{Murlidhar2014}.
Even if uniform charge density is intended for a surface, some variation in charge is unavoidable, especially given the difficulty in controlling precisely  the surface chemistry producing the charge, and this will give rise to non-uniform electro-osmotic flows even in a straight channel.  One of the conceptually simpler non-uniform electroosmotic flows that can be produced is generated by a sinusoidally periodic surface charge on each side of a straight channel \cite{Ajdari1995}; see Fig.\ref{Ajdari}.  This charge pattern leads to a spatially periodic charge in the double layer adjacent to the boundaries. When a voltage is applied along the channel, the velocity near the surface varies periodically as well, and acts like a periodic 'slip' velocity along the surface, generating a complex cellular flow in the fluid in the channel.  This flow is attractive as a simple boundary condition (straight walls, periodic charge) that nevertheless generates a complex flow, and that, moreover, has an analytic solution in the limit of creeping flow and thin double layers \cite{Ajdari1995}. It is easy to add a uniform surface charge density to the periodic charge, mimicking, for example, an imperfectly treated surface with charge non-uniformity. A periodic deviation from a uniform charge might produce a deviation in the mean flow rate in the channel owing to nonlinear coupling between the flow produced by the uniform charge, and that produced by the periodic charge variations.  If the wall charge varies sinusoidally around zero, the electro-osmotic flow that is generated is periodic, and in the Stokes flow limit has no net flow direction. 

Here we consider the effect of inertia on this simple flow. We employ a spectral method to solve the two-dimensional Navier-Stokes equations, and find, surprisingly, for the case of zero average surface charge, so that the Stokes flow is periodic with no net flow, that at a modest Reynolds number $Re=v_0 L/\nu$ of around $20$, there is a bifurcation to a secondary flow with a non-zero mean flow, even when there is no mean flow induced by the boundary conditions themselves.  Here, $v_0$ is the characteristic velocity, $L$ is the half-width of the channel, and $\nu$ is the kinematic viscosity of the fluid. The presence of this bifurcation means that, even for a boundary condition with no mean surface charge, and hence no mean current, a rectified mean flow can be produced through a purely oscillatory boundary condition.  The direction of the mean flow, to the right or the left, is arbitrary, but could be imposed by adding some small bias to the initial oscillatory flow, either electrically, geometrically, or in some other way. We believe that this is the first report of this hydrodynamic instability in a periodically patterned electroosmotic flow (although this discovery was alluded to in an earlier co-authored by one of the present authors \cite{Zhao2013}).  This bifurcation is of interest in its own right, but might also be a means of generating rectified flow in a channel with no net imposed current.  In fact, since the base flow is completely periodic, the applied voltage along the channel could in principle also be alternating, without changing either the zero net current, or the direction of the resulting flow. To reach the bifurcation condition with a wall charge that varies with position sinusoidally around zero, the Reynolds number must reach a value of close to $20$, which, for water with $\nu = 10^{-6}$m$^2/$s, requires a flow velocity and channel width $2L$ that are relatively large. The flow velocity $v_0$ is given by around $\sigma_0 E/\mu\kappa$, where $\sigma_0$ is the amplitude of the surface charge density, $E$ the electric field imposed parallel to the walls, $\mu$ the fluid viscosity and $\kappa$ the inverse Debye length at the wall \cite{Ajdari1995}, where typical values are $\sigma_0 \sim 1$ charge/nm$^{2}$, $\mu = 10^{-3}$Pa\,s, and $\kappa^{-1} = 10$nm. Under these conditions, a field of $10^4$V/m would yield a velocity of $10^{-2}$m/s, and so a channel of width $2L=2$mm would suffice to yield an instability.  Note that the channel depth would need also to be comparable, or larger than, this scale, to prevent viscous suppression of the instability. 

This instability is distinct from well-known electrokinetic flow instabilities that result from coupling of electric fields and ionic conductivity gradients \cite{Oddy2001, Posner2006, Lin2009, Posner2012}, since there are no ionic conductivity gradients considered in what we report here. Such gradients can arise when a core fluid flow is focused by a 'sheath' fluid introduced at the walls of a microfluidic device, if the two fluids have differing ionic strengths. An instability occurs in such flows at a critical electric current Rayleigh number of $Ra_e = (\epsilon E_a^2 d^2/D\mu)((\gamma-1)/\gamma) |\text{grad}^*\sigma^*|_{\text{max}} = 205$, as reported by Posner and Santiago \cite{Posner2006} and Posner \emph{et al.} \cite{Posner2012}.  Here $\epsilon$ is the fluid permittivity, $E_a$ is the applied electric field, $d$ is the channel depth, $D$ is the ion diffusivity, $\mu$ is the fluid viscosity, $\gamma$ is the ratio of core-to-sheath conductivities, and $|\text{grad}^*\sigma^*|_{\text{max}}$ is the maximum dimensionless conductivity gradient.  Note that the Rayleigh number representing the driving force for this instability disappears when the conductivities of the two fluids are equal to each other (i.e., $\gamma = 1$), since then there is no conductivity gradient to which the electric field can couple and the above Rayleigh number is zero. If a conductivity gradient is present,  the instability produced by it could in many cases occur at lower field strength than that needed to produce the inertial instability to be discussed here.  

We consider the flow of a Newtonian fluid in a $2D$-channel forced by a prescribed slip velocity at the channel walls, similar to Fig.\ref{Ajdari}. We introduce a Cartesian coordinate system with the $x$-axis pointing along the length of the channel, and the $y$-axis -- in the wall-normal direction. The walls are located at $y=\pm L$, so that the total channel width is $2L$. The velocity components are ${\bf v}=(u(x,y,t),v(x,y,t))$, and the flow is driven by the following slip velocity at the walls
\begin{align}
& u(x,L,t) = v_+ \cos{kx}, \nonumber \\
& u(x,-L,t) = v_- \cos{\left(kx+\phi\right)}, \label{slip} \\
& v(x,\pm L,t) = 0. \nonumber
\end{align}
Here, $v_+$ and $v_-$ are the maximum slip velocities at the corresponding boundary, $k$ is the wavenumber of the slip velocity modulation, and $\phi$ is the phase difference between the velocity at the upper and lower walls. The equations of motion are given by the Navier-Stokes equation
\begin{equation}
\rho \left[ \frac{\partial {\bf v}}{\partial t} + {\bf v}\cdot{\bf \nabla} {\bf v} \right] = -\nabla p + \mu \nabla^2 {\bf v},
\label{NS}
\end{equation}
and the incompressibility condition
\begin{equation}
\nabla\cdot{\bf v} = 0.
\end{equation}
Here, $\rho$ and $\mu$ are the density and viscosity of the fluid, respectively, and $p$ is the pressure. 
The problem is rendered dimensionless by the rescaling of all the variables, where 
we use the half-width of the channel $L$ as the unit of length, $\text{max}(v_+,v_-)$ as the unit of velocity, and $L/\text{max}(v_+,v_-)$ as the unit of time. We also introduce the Reynolds number 
\begin{equation}
Re=\frac{\text{max}(v_0,v_1) L}{\nu},
\end{equation}
and the dimensionless wave-vector $\tilde{k} = k L$. In what follows, all variables are dimensionless unless stated otherwise. 

To reduce the number of degrees of freedom, we introduce the streamfunction $\Psi=\Psi(x,y,t)$, such that
\begin{equation}
u = \frac{\partial\Psi}{\partial y}, \qquad v = -\frac{\partial\Psi}{\partial x}.
\label{vel_stream}
\end{equation}
In terms of the streamfunction, the equation of motion is given by
\begin{equation}
\left( \frac{\partial}{\partial t} + \frac{\partial \Psi}{\partial y}\frac{\partial}{\partial x} - \frac{\partial \Psi}{\partial x}\frac{\partial}{\partial y} \right)\nabla^2 \Psi = \frac{1}{Re} \nabla^4 \Psi,
\label{eomdim}
\end{equation}
with the following boundary conditions
\begin{align}
&\frac{\partial \Psi}{\partial y}(x,1,t) = \tilde{v}_+ \cos{\tilde{k} x}, \nonumber\\
&\frac{\partial \Psi}{\partial y}(x,-1,t) = \tilde{v}_-  \cos{\left(\tilde{k} x+\phi\right)},\label{bc2dim}\\
&\frac{\partial \Psi}{\partial x}(x,\pm 1,t) = 0, \nonumber
\end{align}
where $\tilde{v}_+=v_+/\text{max}(v_+,v_-)$, and $\tilde{v}_-=v_-/\text{max}(v_+,v_-)$, and $\nabla^4$ is the biharmonic operator.

\begin{figure}
\includegraphics[width=0.9\textwidth]{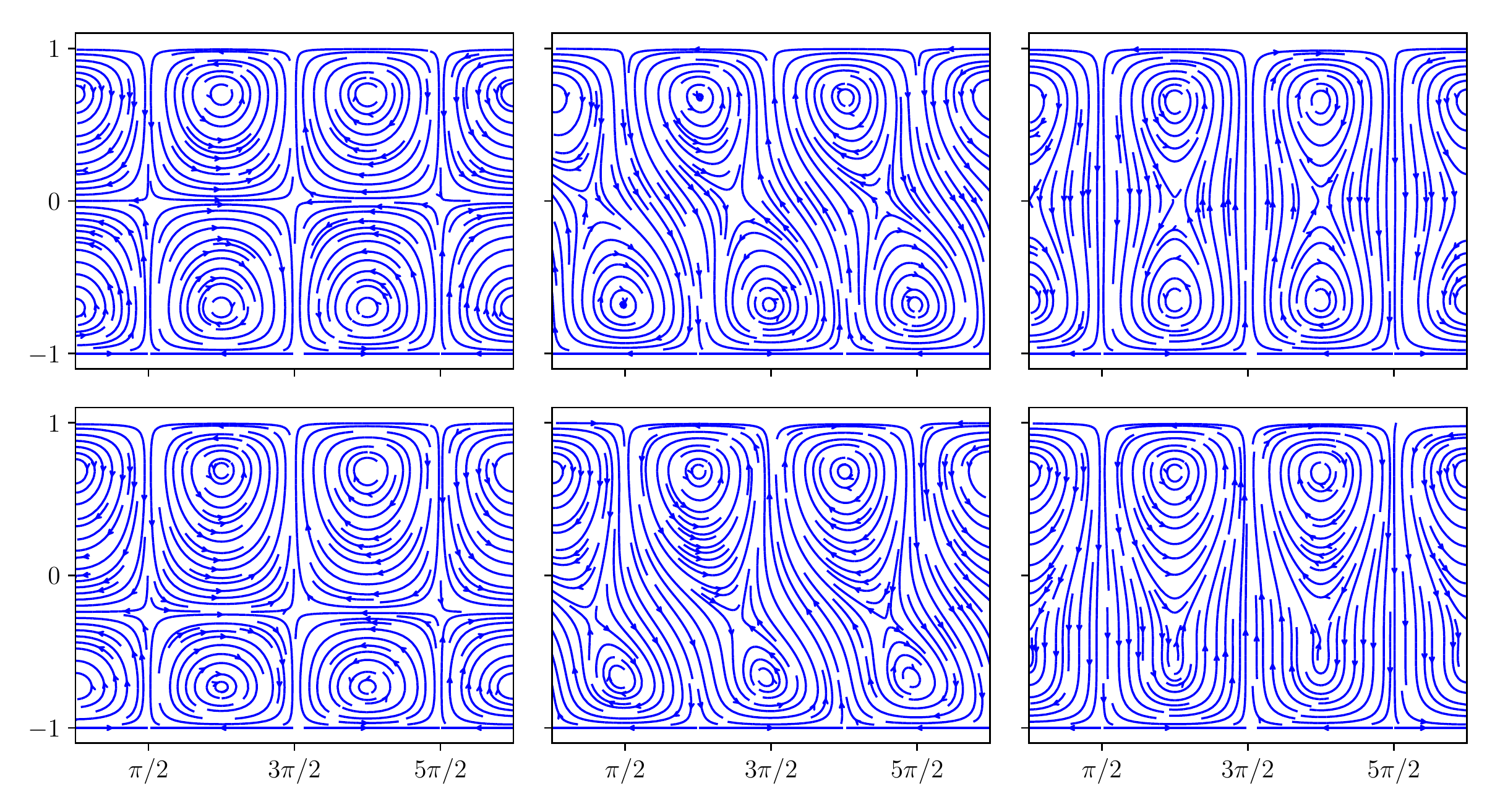}
\caption{Velocity profiles given by Eq.\eqref{stokes_stream} for $\tilde{k}=\pi$. Vertical axes are positions in the gap ($y$-coordinates), and the horizontal axes give the distance along the channel in units of $\tilde{k} x$. The top row corresponds to the symmetric boundary conditions ($\tilde{v}_+=\tilde{v}_-=1$), while the bottom row shows the effect of their asymmetry ($\tilde{v}_+=1$ and $\tilde{v}_-=1/3$). The phase shift $\phi$ is: (left) $0$, (middle) $\pi/2$, and (right) $\pi$. }
\label{stokes_flow}
\end{figure}

It is instructive to consider the limit of zero inertia and steady velocity field. In this case, Eq.(\ref{eomdim}) reduces to the Stokes equation, $\nabla^4 \Psi = 0$, that has a simple analytical solution satisfying the boundary conditions, Eqs.\eqref{bc2dim},
\begin{align}
&\Psi_0 = \frac{2\tilde{k}}{\sinh^2{2\tilde{k}}-4\tilde{k}^2}\biggl[
A(x)\left(1+y\right)\sinh{\tilde{k}\left(1-y\right)}  \nonumber\\
&\quad\qquad \qquad \qquad \qquad \qquad \quad - B(x)\left(1-y\right)\sinh{\tilde{k}\left(1+y\right)} \biggr],
\label{stokes_stream}
\end{align}
where
\begin{eqnarray}
A(x) = \tilde{v}_+ \cos{\tilde{k}x} + \tilde{v}_- \frac{\sinh{2\tilde{k}}}{2\tilde{k}}  \cos{\left(\tilde{k}x+\phi\right)}, \\
B(x) = \tilde{v}_- \cos{\left(\tilde{k}x+\phi\right)} + \tilde{v}_+ \frac{\sinh{2\tilde{k}}}{2\tilde{k}}  \cos{\tilde{k}x}.
\end{eqnarray}
This solution is similar to the one obtained by Ajdari \cite{Ajdari1995}. Although $\Psi_0$ differs from the true solution to Eq.\eqref{eomdim} for any finite amount of inertia, it is nevertheless useful for gaining an insight into the structure of the flow.  In Fig.\ref{stokes_flow} we plot the velocity profile given by $\Psi_0$ for $\tilde{v}_+=\tilde{v}_-=1$ (top row), and $\tilde{v}_+=1$, $\tilde{v}_-=1/3$ (bottom row) for three values of the phase difference $\phi$: $0$ (left column), $\pi/2$ (middle column), and $\pi$ (right column). As can be seen from the figure, the flow consists of two arrays of vortices aligned along each wall with their relative position and strength set by the phase-difference $\phi$ and the velocity magnitudes $\tilde{v}_+$ and $\tilde{v}_-$, respectively.

To assess the effect of inertia on this solution, we solve Eqs.\eqref{eomdim} and \eqref{bc2dim} numerically using a Fourier-Chebyshev pseudo-spectral method \cite{boydbook,canutobook}. We express the streamfunction as a Fourier series
\begin{equation}
\Psi(x,y,t) = \sum_{n=-N}^{N} \psi_{n}(y,t)e^{i n \tilde{k} x}, 
\end{equation}
where $\psi_{n}(y,t) = \psi_{-n}^{*}(y,t)$ to ensure that $\Psi(x,y,t)$ is real, and $*$ denotes the complex conjugate. At any time $t$, $\psi_{n}(y,t)$ is represented by its values at $M$ Gauss-Lobatto points \cite{canutobook} in the wall-normal direction, and the $y$-derivatives are taken by multiplying these values with the Chebyshev pseudo-spectral differentiation matrix \cite{canutobook}. The non-linear terms are calculated by performing a discrete Fourier transform of the streamfunction to real space, evaluating the non-linear terms there, and performing an inverse discrete Fourier transform back to spectral space; the $3/2$-rule is used to avoid aliasing errors and the boundary conditions are implemented using the tau-method \cite{boydbook,canutobook}. For each set of parameters, we check convergence of the velocity field by comparing it at several resolutions $(N,M)$; convergence was always reached for $N=5$ (before de-aliasing) and $M=80$. Most of the results presented below are obtained by using the Newton-Raphson algorithm \cite{canutobook} to solve the time-independent version of Eq.\eqref{eomdim}. We also performed direct numerical simulations of Eq.\eqref{eomdim} using a fully-implicit Crank-Nicolson method \cite{boydbook,canutobook}; for all parameters studied, convergence was reached for the dimensionless time-step of $10^{-2}$.

\begin{figure}[t]
\includegraphics[width=0.5\textwidth]{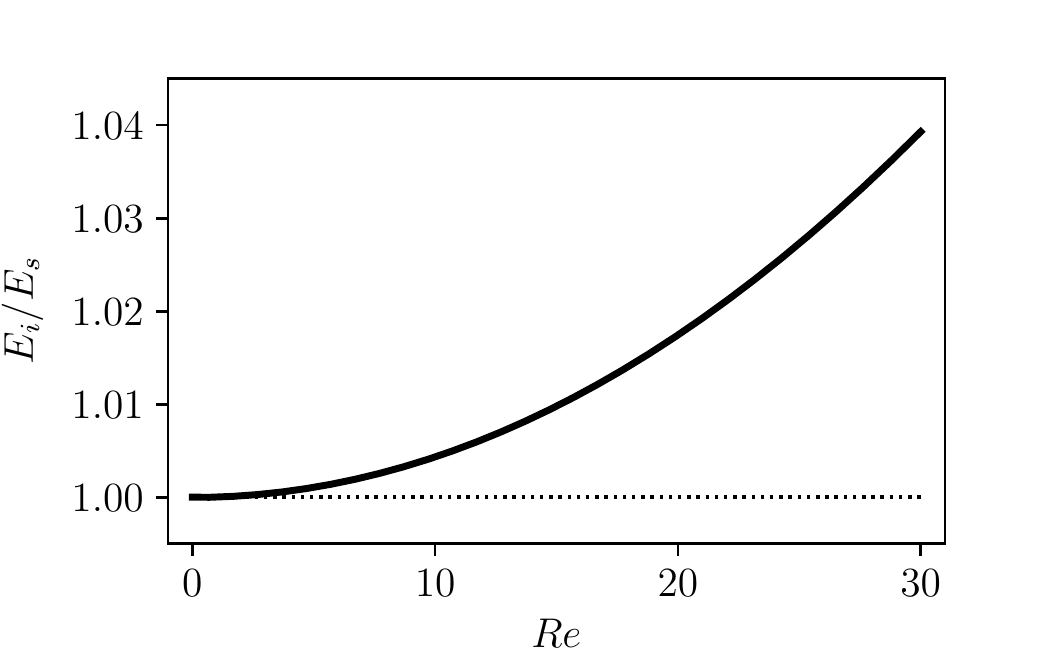}
\caption{The ratio of the kinetic energies of the inertial and Stokes solutions for $\tilde{v}_+=\tilde{v}_-=1$, $\phi=0$, and $\tilde{k}=\pi$ as a function of the Reynolds number $Re$. The solid line is well-approximated by $1+(Re/151.41)^2$.}
\label{Ecomp}
\end{figure}

First, we study how the presence of inertia modifies the Stokes solution, Eq.\eqref{stokes_stream}, at relatively low Reynolds numbers. Using the Newton-Raphson method, we find steady solutions of Eq.\eqref{eomdim}, and compare them to the Stokes profile $\Psi_0$. The difference is quantified by calculating the kinetic energy of the flow, defined as
\begin{equation}
E = \frac{\tilde{k}}{2\pi}\int_0^{\frac{2\pi}{\tilde{k}}}dx\,\,\frac{1}{2}\int_{-1}^{1}dy \frac{1}{2}\left[ \left(\frac{\partial\Psi}{\partial y}\right)^2+\left(\frac{\partial\Psi}{\partial x}\right)^2 \right],
\end{equation}
for the inertial, $E_i$, and Stokes solutions, $E_s$. In Fig.\ref{Ecomp} we plot the ratio $E_i/E_s$ for $\tilde{v}_+=\tilde{v}_-=1$, $\phi=0$, and $\tilde{k}=\pi$. The data demonstrate that the inertial contribution to the kinetic energy is only about $4\%$ of the total kinetic energy at $Re=30$, and that that contribution decreases for smaller values of $Re$.  The difference $E_i/E_s-1$  scales quadratically with $Re$ (fit not shown), implying that the small inertial correction to the Stokes profile can be obtained from the leading-order term of the perturbation theory in $Re$, even for $Re\sim 30$. Visual inspection of the inertial velocity profiles together with the data in Fig.\ref{Ecomp} suggests that the Stokes solution, Eq.\eqref{stokes_stream}, is a very good approximation to the actual inertial solution even at moderate Reynolds numbers. 

\begin{figure}[t]
\includegraphics[width=0.49\textwidth]{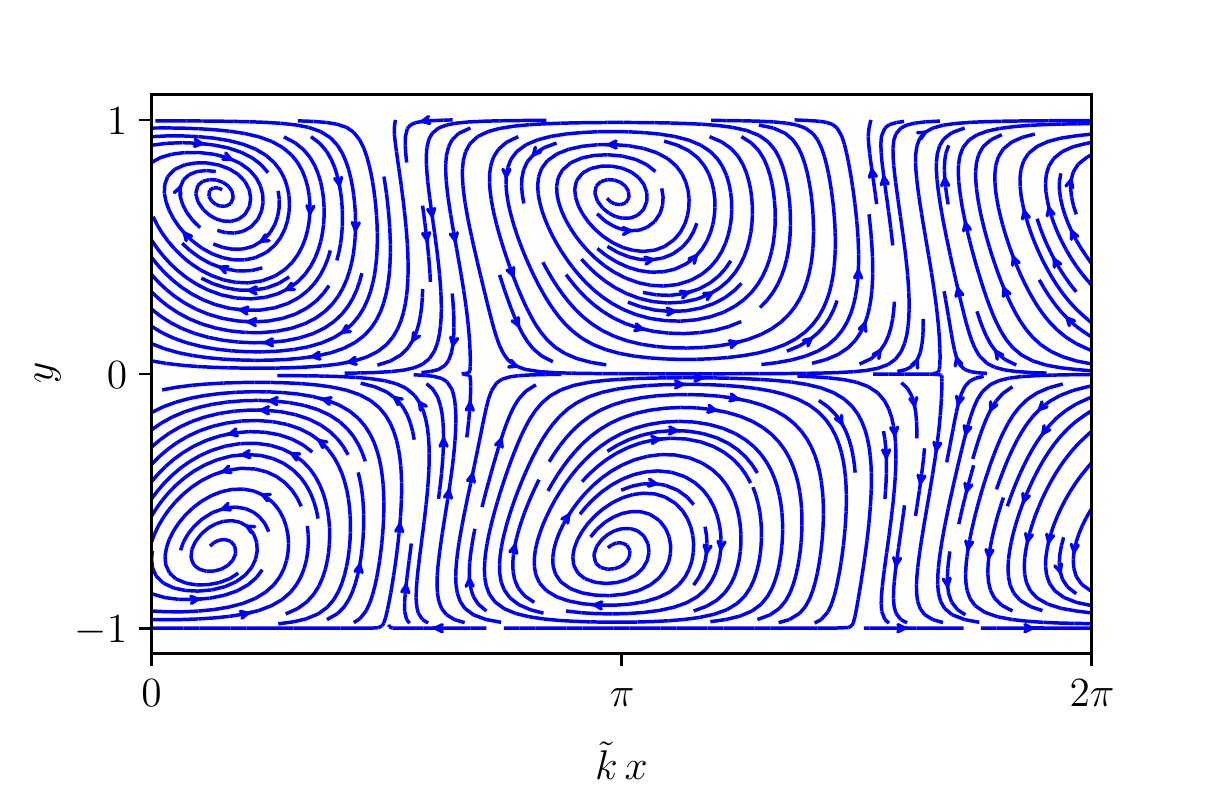}
\includegraphics[width=0.49\textwidth]{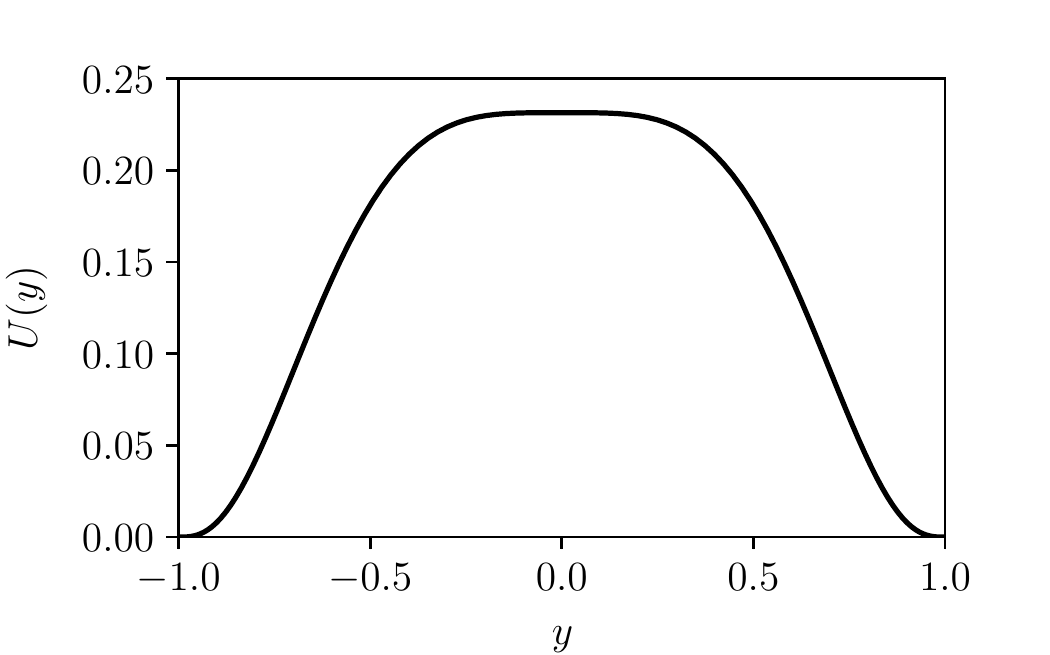}
\caption{Velocity profile at $Re=40$ for $\tilde{v}_+=\tilde{v}_-=1$,  $\phi=0$, and $\tilde{k}=\pi$. (Left) The velocity field in the channel without its mean profile (zeroth Fourier harmonics). (Right) The mean profile $U(y)$.}
\label{bifflow}
\end{figure}

The situation changes significantly at higher Reynolds numbers. In Fig.\ref{bifflow}(left) we plot the velocity profile for the symmetric boundary conditions at $Re=40$ and $\tilde{k}=\pi$, and observe that it no longer posseses a translation-reflection symmetry along the $x$-axis, c.f. Fig.\ref{stokes_flow}(top, left). This is associated with the emergence of the zeroth Fourier mode $U(y)$ of the horizontal velocity component $u(x,y)$, see Fig.\ref{bifflow}(right), absent at lower Reynolds numbers. This $x$-independent, mean flow along the $x$-direction reaches significant amplitudes of about $23\%$ of the maximum slip velocity at the wall.

\begin{figure}[t]
\includegraphics[width=0.49\textwidth]{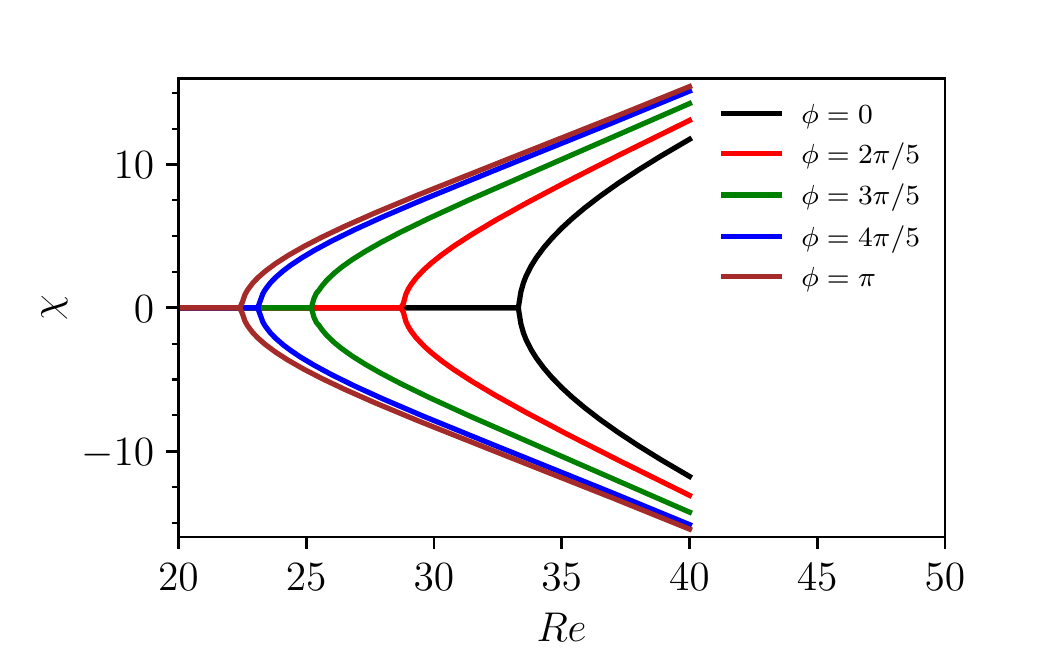}
\includegraphics[width=0.5\textwidth]{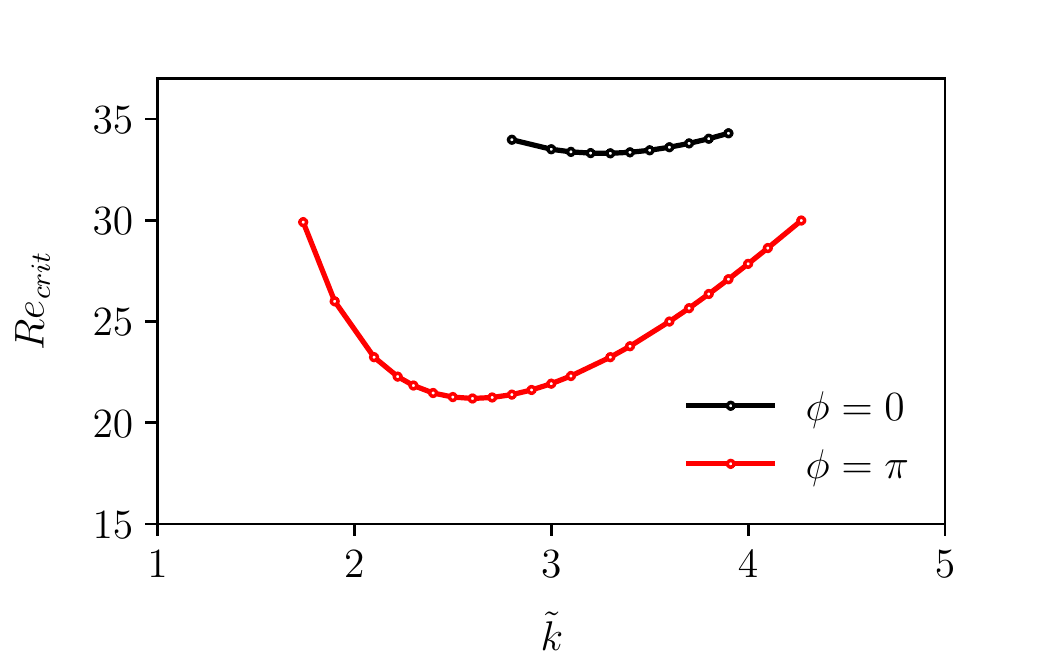}
\caption{(Left) The bifurcation diagram, $\chi$ vs $Re$, for various values of $\phi$ at $\tilde{k}=\pi$. (Right) The critical Reynolds number $Re_{crit}$ as a function of $\tilde{k}$. For $\phi=0$ the minimal value of $Re_{crit}$ is $Re_{crit}^{min}=33.31$ at $\tilde{k}=3.3$, while for $\phi=\pi$,  $Re_{crit}^{min}=21.2$ at $\tilde{k}=2.6$. In both plots $\tilde{v}_+=\tilde{v}_-=1$.}
\label{bifurc}
\end{figure}

To characterise this new flow state, we introduce a dimensionless order parameter
\begin{equation}
\chi = Re \frac{\tilde{k}}{2\pi}\int_0^{\frac{2\pi}{\tilde{k}}} dx \int_{-1}^{1} dy\,u(x,y) \equiv
Re \int_{-1}^{1} U(y) dy,
\end{equation}
which is a two-dimensional flow rate along the channel (in physical units) scaled by the kinematic viscosity $\nu$ of the fluid. In Fig.\ref{bifurc}(left) we plot $\chi$ as a function of the Reynolds number for $\tilde{v}_+=\tilde{v}_-=1$, $\phi=0$, and $\tilde{k}=\pi$ (black line). For low values of $Re$ the flow is left-right-symmetric, there is no mean flow, and $\chi=0$, while at larger $Re$, $\chi$ acquires non-zero values indicating the presence of a mean flow. The direction of the mean flow is selected by a spontaneous symmetry breaking, and can be in either direction along the channel. The state diagram, Fig.\ref{bifurc}(left), therefore has two symmetric branches, $\pm\chi$, typical of a super-critical (pitchfork) bifurcation. By combining the Newton-Raphson and time-iteration techniques, we have verified that the left-right symmetric solution with $\chi=0$ is also present for higher values of $Re$ but is linearly unstable. The final flow state with $\chi \ne 0$ is stationary and stable with respect to small perturbations. Therefore, we conclude that the new flow state is a result of a linear instability that sets in at $Re_{crit}\approx 33.3$, for this set of parameters. In Fig.\ref{bifurc}(left) we also show the bifurcation diagrams for other values of the phase-difference $\phi$, and observe that the lowest $Re_{crit}$ is achieved for $\phi=\pi$; the corresponding base profile is shown in Fig.\ref{stokes_flow}(top,right). 

The instability thresholds presented above were calculated by imposing a fixed value of $\tilde{k}$, i.e. assuming a particular spatial period of the solution. To find the critical condition in an infinitely long channel, we now study how $Re_{crit}$ depends on $\tilde{k}$. In Fig.\ref{bifurc}(right) we plot the non-linear stability thresholds for two values of $\phi$, and observe that $Re_{crit}^{min} = 21.2$ for $\tilde{k}=2.6$ and $\phi=\pi$. The stability thresholds for other values of $\phi$ lie in-between the two cases presented in Fig.\ref{bifurc}(right), similar to Fig.\ref{bifurc}(left).

We also studied the effect of the asymmetry in the wall slip velocity (not shown), with either $\tilde{v}_+$ or $\tilde{v}_-$ smaller than unity. For every set of $\phi$ and $\tilde{k}$ considered, the corresponding $Re_{crit}$ was found to be larger than $Re_{crit}$ for $\tilde{v}_+=\tilde{v}_-=1$.

As mentioned in the Introduction, this instability can potentially be utilised as a means of creating a unidirectional flow in a microfluidic device, although relatively high transitional Reynolds numbers and the a priory unknown direction of the flow could make it impractical. We now attempt to assess whether a modification of the slip boundary condition, Eq.\eqref{slip}, can alleviate both problems. To this end, we consider the following (dimensional) velocity profile prescribed at the walls
\begin{align}
u(x,\pm L,t) = v_c \pm v \cos{kx}, \label{slip_plug}
\end{align}
where the spatially-oscillatory component is the same as in Eq.\eqref{slip} for the most unstable parameters ($v_+=v_-\equiv v$, $\phi=\pi$, and $k L=2.6$), and we have introduced $v_c$ -- the amplitude of a constant slip velocity in the positive $x$-direction. Here, we study whether a small value of $v_c$ can produce a significant mean flow at $Re<Re_{crit}^{min}$. The problem is made dimensionless as before, and we define additionally another Reynolds number, $Re_c=v_c L/\nu$, based on the constant slip velocity. 

\begin{figure}[t]
\includegraphics[width=0.5\textwidth]{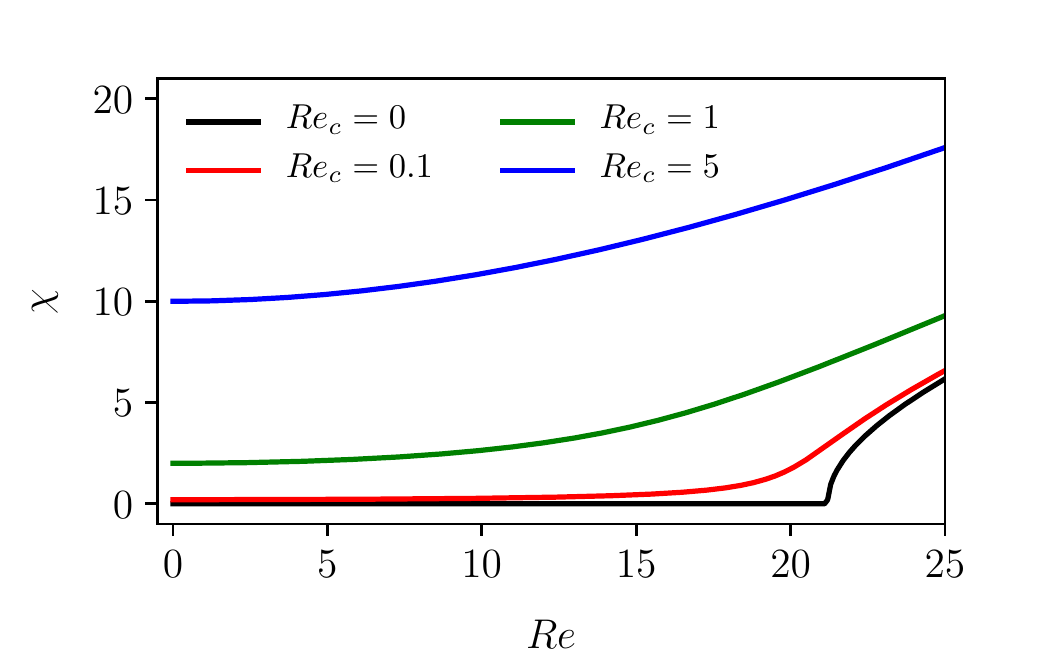}
\caption{The bifurcation diagram for the modified boundary conditions with a bias, Eq.\eqref{slip_plug} for $\phi=\pi$ and $\tilde{k}=2.6$. The strength of the bias, $Re_c$, is fixed and the strength of the spatially oscillatory component, $Re$, is varied.}
\label{pump}
\end{figure}

In the absence of the spatially-oscillatory component, equations of motion are trivially solved by a plug-like flow, $U(y)=v_c$ (in physical units),which to the order parameter given by $\chi_c=2Re_c$. For $Re>0$, we expect that the interaction between the plug-like and spatially-oscillatory components will generate flow rates enhanced beyond $\chi_c$. In Fig.\ref{pump} we present the bifurcation diagram for the modified boundary conditions, Eq.\eqref{slip_plug}, varying $Re$ but keeping $Re_c$ fixed to a particular value. The $Re_c=0$ data is the same as in Fig.\ref{bifurc}(left) for $\phi=\pi$. In the presence of a constant bias, the bifurcation diagram loses its $\pm\chi$ symmetry, and we only plot the dimensionless flow rate  in the same direction as the bias. For $Re_c=0.1$ and $Re_c=1$, the flow rate is dominated by the plug-like profile at low values of $Re$, while at larger $Re$ there is an enhancement of the flow rate due to the instability. The bifurcation diagram now looks like an imperfect pitch-fork bifurcation. For yet larger $Re_c$, the effect of the underlying instability is masked by the presence of a strong bias and only a mild enhancement is observed. While the presence of the bias clearly enhances the mean flow rate and breaks the left-right symmetry, the enhancement is mild and it remains to be seen whether there are practical advantages of generating a steady flow in a microfluidic device by a slip velocity Eq.\eqref{slip_plug} instead of a stronger steady component alone.

The results developed here have several implications.  First, we find that a periodic variation in wall charge will have only a small effect on average velocity in a microfluidic device with an otherwise uniform wall charge, if the periodic component is small (or even modest) in magnitude compared to the uniform component. This conclusion seems likely to hold if the non-uniform component is irregular or non-periodic, as long as it is significantly smaller than the uniform wall charge.  Thus, surface charge in a microfluidic device does not need to be nearly perfectly uniform to achieve a uniform flow rate, whose magnitude is set by the average surface charge, a conclusion of importance in practical applications where wall charging is unlikely to be exquisitely uniform.  Secondly, if a rectified flow with a sharp onset is desired in a microfluidic device using electric fields to drive the flow,  this can be accomplished by exploiting the bifurcation described here, albeit only for rather large channel widths and heights (i.e., millimeters) and strong fields.  In addition, there may be benefit in using periodic, or nearly periodic flows for separation of particles or cells based on size or other characteristics, including separations based on inertial forces.  These inertial forces are already being exploited in pressure-driven flows to separate rare circulating tumor cells from white blood cells \cite{Murlidhar2014}.  Electroosmotic flow driven by a periodic wall charge, along with fluid inertial forces, may expand the options for improving the efficiency of such devices. We note that inertial fluid forces in pressure-driven microfluidic devices are strong enough to induced circulating Dean flows, which are of great significance for separating particles and cells. Thus, the addition of electroosmotically driven flow, combined with inertial effects, opens multiple new opportunities for separations. Thirdly, the flows generated by periodic charges may provide a good experimental test of one's ability to control elecroosmotic flow fields, and of the ability to created controlled charge at walls.  Since the flow field is readily predicted, including the effect of surface charge amplitude and other parameters,  a measurement of the flow (even without the bifurcation) could be used to validate methods of controlling surface charge, for example. Fourthly, both the circulating primary flow and the secondary bifurcation flow described here occurs in a geometry of trivial simplicity  (a straight channel), which allows it to be used as a test flow field for exploring various advanced simulating methods, such as mesoscopic flow simulations \cite{Zhao2013}, and for exploring the behaviour of complex fluids in complex flows, but with simple geometry and boundary conditions \cite{Rezvantalab2016}. Finally, the flow is essentially completely viscous prior to the bifurcation and described by an analytical solution to the Stokes equation, and thus represents a particularly simple and elegant example of a classical forward bifurcation at a very modest Reynolds number, and is the simplest bifurcation so far presented for electroosmotic flow.  

R.G.L. acknowledges support and hospitality of the Higgs Centre for Theoretical Physics, University of Edinburgh where a part of this work was performed. A.M. acknowledges support from the UK Engineering and Physical Sciences Research Council (EP/I004262/1). Research outputs generated through the EPSRC grant EP/I004262/1 can be found at http://dx.doi.org/xxx-xxx.

\bibliography{lit}

\begin{thebibliography}{15}%
\makeatletter
\providecommand \@ifxundefined [1]{%
 \@ifx{#1\undefined}
}%
\providecommand \@ifnum [1]{%
 \ifnum #1\expandafter \@firstoftwo
 \else \expandafter \@secondoftwo
 \fi
}%
\providecommand \@ifx [1]{%
 \ifx #1\expandafter \@firstoftwo
 \else \expandafter \@secondoftwo
 \fi
}%
\providecommand \natexlab [1]{#1}%
\providecommand \enquote  [1]{``#1''}%
\providecommand \bibnamefont  [1]{#1}%
\providecommand \bibfnamefont [1]{#1}%
\providecommand \citenamefont [1]{#1}%
\providecommand \href@noop [0]{\@secondoftwo}%
\providecommand \href [0]{\begingroup \@sanitize@url \@href}%
\providecommand \@href[1]{\@@startlink{#1}\@@href}%
\providecommand \@@href[1]{\endgroup#1\@@endlink}%
\providecommand \@sanitize@url [0]{\catcode `\\12\catcode `\$12\catcode
  `\&12\catcode `\#12\catcode `\^12\catcode `\_12\catcode `\%12\relax}%
\providecommand \@@startlink[1]{}%
\providecommand \@@endlink[0]{}%
\providecommand \url  [0]{\begingroup\@sanitize@url \@url }%
\providecommand \@url [1]{\endgroup\@href {#1}{\urlprefix }}%
\providecommand \urlprefix  [0]{URL }%
\providecommand \Eprint [0]{\href }%
\providecommand \doibase [0]{http://dx.doi.org/}%
\providecommand \selectlanguage [0]{\@gobble}%
\providecommand \bibinfo  [0]{\@secondoftwo}%
\providecommand \bibfield  [0]{\@secondoftwo}%
\providecommand \translation [1]{[#1]}%
\providecommand \BibitemOpen [0]{}%
\providecommand \bibitemStop [0]{}%
\providecommand \bibitemNoStop [0]{.\EOS\space}%
\providecommand \EOS [0]{\spacefactor3000\relax}%
\providecommand \BibitemShut  [1]{\csname bibitem#1\endcsname}%
\let\auto@bib@innerbib\@empty
\bibitem [{\citenamefont {Pretorius}\ \emph {et~al.}(1974)\citenamefont
  {Pretorius}, \citenamefont {Hopkins},\ and\ \citenamefont
  {Schieke}}]{Pretorius1974}%
  \BibitemOpen
  \bibfield  {author} {\bibinfo {author} {\bibfnamefont {V.}~\bibnamefont
  {Pretorius}}, \bibinfo {author} {\bibfnamefont {B.~J.}\ \bibnamefont
  {Hopkins}}, \ and\ \bibinfo {author} {\bibfnamefont {J.~D.}\ \bibnamefont
  {Schieke}},\ }\href@noop {} {\bibfield  {journal} {\bibinfo  {journal} {J.
  Chromatogr. A}\ }\textbf {\bibinfo {volume} {99}},\ \bibinfo {pages} {23}
  (\bibinfo {year} {1974})}\BibitemShut {NoStop}%
\bibitem [{\citenamefont {Probstein}(1994)}]{Probstein1994}%
  \BibitemOpen
  \bibfield  {author} {\bibinfo {author} {\bibfnamefont {R.}~\bibnamefont
  {Probstein}},\ }\href@noop {} {\emph {\bibinfo {title} {{Physicochemical
  Hydrodynamics}}}}\ (\bibinfo  {publisher} {Wiley, New York},\ \bibinfo {year}
  {1994})\BibitemShut {NoStop}%
\bibitem [{\citenamefont {Schoch}\ \emph {et~al.}(2008)\citenamefont {Schoch},
  \citenamefont {Han},\ and\ \citenamefont {Renaud}}]{Schoch2008}%
  \BibitemOpen
  \bibfield  {author} {\bibinfo {author} {\bibfnamefont {R.~B.}\ \bibnamefont
  {Schoch}}, \bibinfo {author} {\bibfnamefont {J.}~\bibnamefont {Han}}, \ and\
  \bibinfo {author} {\bibfnamefont {P.}~\bibnamefont {Renaud}},\ }\href@noop {}
  {\bibfield  {journal} {\bibinfo  {journal} {Rev. Mod. Phys.}\ }\textbf
  {\bibinfo {volume} {80}},\ \bibinfo {pages} {839} (\bibinfo {year}
  {2008})}\BibitemShut {NoStop}%
\bibitem [{\citenamefont {Wang}\ \emph {et~al.}(2009)\citenamefont {Wang},
  \citenamefont {Cheng}, \citenamefont {Wang},\ and\ \citenamefont
  {Liu}}]{Wang2009}%
  \BibitemOpen
  \bibfield  {author} {\bibinfo {author} {\bibfnamefont {X.}~\bibnamefont
  {Wang}}, \bibinfo {author} {\bibfnamefont {C.}~\bibnamefont {Cheng}},
  \bibinfo {author} {\bibfnamefont {S.}~\bibnamefont {Wang}}, \ and\ \bibinfo
  {author} {\bibfnamefont {S.}~\bibnamefont {Liu}},\ }\href@noop {} {\bibfield
  {journal} {\bibinfo  {journal} {Microfluid. Nanofluid.}\ }\textbf {\bibinfo
  {volume} {6}},\ \bibinfo {pages} {145–162} (\bibinfo {year}
  {2009})}\BibitemShut {NoStop}%
\bibitem [{\citenamefont {Lee}\ \emph {et~al.}(2011)\citenamefont {Lee},
  \citenamefont {Chang}, \citenamefont {Wang},\ and\ \citenamefont
  {Fu}}]{Lee2011}%
  \BibitemOpen
  \bibfield  {author} {\bibinfo {author} {\bibfnamefont {C.-Y.}\ \bibnamefont
  {Lee}}, \bibinfo {author} {\bibfnamefont {C.-L.}\ \bibnamefont {Chang}},
  \bibinfo {author} {\bibfnamefont {Y.-N.}\ \bibnamefont {Wang}}, \ and\
  \bibinfo {author} {\bibfnamefont {L.-M.}\ \bibnamefont {Fu}},\ }\href@noop {}
  {\bibfield  {journal} {\bibinfo  {journal} {Int. J. Mol. Sci.}\ }\textbf
  {\bibinfo {volume} {12}},\ \bibinfo {pages} {3263} (\bibinfo {year}
  {2011})}\BibitemShut {NoStop}%
\bibitem [{\citenamefont {Murlidhar}\ \emph {et~al.}(2014)\citenamefont
  {Murlidhar}, \citenamefont {Zeinali}, \citenamefont {Grabauskiene},
  \citenamefont {Ghannad-Rezaie}, \citenamefont {Wicha}, \citenamefont
  {Simeone}, \citenamefont {Ramnath}, \citenamefont {Reddy},\ and\
  \citenamefont {Nagrath}}]{Murlidhar2014}%
  \BibitemOpen
  \bibfield  {author} {\bibinfo {author} {\bibfnamefont {V.}~\bibnamefont
  {Murlidhar}}, \bibinfo {author} {\bibfnamefont {M.}~\bibnamefont {Zeinali}},
  \bibinfo {author} {\bibfnamefont {S.}~\bibnamefont {Grabauskiene}}, \bibinfo
  {author} {\bibfnamefont {M.}~\bibnamefont {Ghannad-Rezaie}}, \bibinfo
  {author} {\bibfnamefont {M.~S.}\ \bibnamefont {Wicha}}, \bibinfo {author}
  {\bibfnamefont {D.~M.}\ \bibnamefont {Simeone}}, \bibinfo {author}
  {\bibfnamefont {N.}~\bibnamefont {Ramnath}}, \bibinfo {author} {\bibfnamefont
  {R.~M.}\ \bibnamefont {Reddy}}, \ and\ \bibinfo {author} {\bibfnamefont
  {S.}~\bibnamefont {Nagrath}},\ }\href@noop {} {\bibfield  {journal} {\bibinfo
   {journal} {Small}\ }\textbf {\bibinfo {volume} {10}},\ \bibinfo {pages}
  {4895} (\bibinfo {year} {2014})}\BibitemShut {NoStop}%
\bibitem [{\citenamefont {Ajdari}(1995)}]{Ajdari1995}%
  \BibitemOpen
  \bibfield  {author} {\bibinfo {author} {\bibfnamefont {A.}~\bibnamefont
  {Ajdari}},\ }\href@noop {} {\bibfield  {journal} {\bibinfo  {journal} {Phys.
  Rev. Lett.}\ }\textbf {\bibinfo {volume} {75}},\ \bibinfo {pages} {755}
  (\bibinfo {year} {1995})}\BibitemShut {NoStop}%
\bibitem [{\citenamefont {Zhao}\ \emph {et~al.}(2013)\citenamefont {Zhao},
  \citenamefont {Wang}, \citenamefont {Jiang},\ and\ \citenamefont
  {Larson}}]{Zhao2013}%
  \BibitemOpen
  \bibfield  {author} {\bibinfo {author} {\bibfnamefont {T.}~\bibnamefont
  {Zhao}}, \bibinfo {author} {\bibfnamefont {X.}~\bibnamefont {Wang}}, \bibinfo
  {author} {\bibfnamefont {L.}~\bibnamefont {Jiang}}, \ and\ \bibinfo {author}
  {\bibfnamefont {R.~G.}\ \bibnamefont {Larson}},\ }\href@noop {} {\bibfield
  {journal} {\bibinfo  {journal} {J. Chem. Phys.}\ }\textbf {\bibinfo {volume}
  {139}},\ \bibinfo {pages} {084109} (\bibinfo {year} {2013})}\BibitemShut
  {NoStop}%
\bibitem [{\citenamefont {Oddy}\ \emph {et~al.}(2001)\citenamefont {Oddy},
  \citenamefont {Santiago},\ and\ \citenamefont {Mikkelson}}]{Oddy2001}%
  \BibitemOpen
  \bibfield  {author} {\bibinfo {author} {\bibfnamefont {M.~H.}\ \bibnamefont
  {Oddy}}, \bibinfo {author} {\bibfnamefont {J.~G.}\ \bibnamefont {Santiago}},
  \ and\ \bibinfo {author} {\bibfnamefont {J.~C.}\ \bibnamefont {Mikkelson}},\
  }\href@noop {} {\bibfield  {journal} {\bibinfo  {journal} {Anal. Chem.}\
  }\textbf {\bibinfo {volume} {73}},\ \bibinfo {pages} {5822} (\bibinfo {year}
  {2001})}\BibitemShut {NoStop}%
\bibitem [{\citenamefont {Posner}\ and\ \citenamefont
  {Santiago}(2006)}]{Posner2006}%
  \BibitemOpen
  \bibfield  {author} {\bibinfo {author} {\bibfnamefont {J.~D.}\ \bibnamefont
  {Posner}}\ and\ \bibinfo {author} {\bibfnamefont {J.~G.}\ \bibnamefont
  {Santiago}},\ }\href@noop {} {\bibfield  {journal} {\bibinfo  {journal} {J.
  Fluid Mech.}\ }\textbf {\bibinfo {volume} {555}},\ \bibinfo {pages} {1–42}
  (\bibinfo {year} {2006})}\BibitemShut {NoStop}%
\bibitem [{\citenamefont {Lin}(2009)}]{Lin2009}%
  \BibitemOpen
  \bibfield  {author} {\bibinfo {author} {\bibfnamefont {H.}~\bibnamefont
  {Lin}},\ }\href@noop {} {\bibfield  {journal} {\bibinfo  {journal} {Mech.
  Res. Commun.}\ }\textbf {\bibinfo {volume} {36}},\ \bibinfo {pages} {33–38}
  (\bibinfo {year} {2009})}\BibitemShut {NoStop}%
\bibitem [{\citenamefont {Posner}\ \emph {et~al.}(2012)\citenamefont {Posner},
  \citenamefont {Perez},\ and\ \citenamefont {Santiago}}]{Posner2012}%
  \BibitemOpen
  \bibfield  {author} {\bibinfo {author} {\bibfnamefont {J.~D.}\ \bibnamefont
  {Posner}}, \bibinfo {author} {\bibfnamefont {C.~L.}\ \bibnamefont {Perez}}, \
  and\ \bibinfo {author} {\bibfnamefont {J.~G.}\ \bibnamefont {Santiago}},\
  }\href@noop {} {\bibfield  {journal} {\bibinfo  {journal} {Proc. Natl. Acad.
  Sci. U.S.A.}\ }\textbf {\bibinfo {volume} {109}},\ \bibinfo {pages} {14353}
  (\bibinfo {year} {2012})}\BibitemShut {NoStop}%
\bibitem [{\citenamefont {Boyd}(2013)}]{boydbook}%
  \BibitemOpen
  \bibfield  {author} {\bibinfo {author} {\bibfnamefont {J.~P.}\ \bibnamefont
  {Boyd}},\ }\href {https://books.google.co.uk/books?id=b4TCAgAAQBAJ} {\emph
  {\bibinfo {title} {{Chebyshev and Fourier Spectral Methods: Second Revised
  Edition}}}},\ Dover Books on Mathematics\ (\bibinfo  {publisher} {Dover
  Publications},\ \bibinfo {year} {2013})\BibitemShut {NoStop}%
\bibitem [{\citenamefont {Canuto}\ \emph {et~al.}(1987)\citenamefont {Canuto},
  \citenamefont {Hussaini}, \citenamefont {Quarteroni},\ and\ \citenamefont
  {Zang}}]{canutobook}%
  \BibitemOpen
  \bibfield  {author} {\bibinfo {author} {\bibfnamefont {C.}~\bibnamefont
  {Canuto}}, \bibinfo {author} {\bibfnamefont {M.~Y.}\ \bibnamefont
  {Hussaini}}, \bibinfo {author} {\bibfnamefont {A.}~\bibnamefont
  {Quarteroni}}, \ and\ \bibinfo {author} {\bibfnamefont {T.~A.}\ \bibnamefont
  {Zang}},\ }\href@noop {} {\emph {\bibinfo {title} {{Spectral Methods in Fluid
  Dynamics}}}}\ (\bibinfo  {publisher} {Springer-Verlag},\ \bibinfo {year}
  {1987})\BibitemShut {NoStop}%
\bibitem [{\citenamefont {Rezvantalab}\ \emph {et~al.}(2016)\citenamefont
  {Rezvantalab}, \citenamefont {Zhu},\ and\ \citenamefont
  {Larson}}]{Rezvantalab2016}%
  \BibitemOpen
  \bibfield  {author} {\bibinfo {author} {\bibfnamefont {H.}~\bibnamefont
  {Rezvantalab}}, \bibinfo {author} {\bibfnamefont {G.}~\bibnamefont {Zhu}}, \
  and\ \bibinfo {author} {\bibfnamefont {R.~G.}\ \bibnamefont {Larson}},\
  }\href@noop {} {\bibfield  {journal} {\bibinfo  {journal} {Soft Matter}\
  }\textbf {\bibinfo {volume} {12}},\ \bibinfo {pages} {5883} (\bibinfo {year}
  {2016})}\BibitemShut {NoStop}%
\end{thebibliography}%

\end{document}